\documentclass{IEEElmagForArxiv}

\usepackage[colorlinks,urlcolor=blue,linkcolor=blue,citecolor=blue]{hyperref}

\usepackage[hyphenbreaks]{breakurl}
\usepackage{physics}
\usepackage{color}
\usepackage{array}

\usepackage{amssymb,amsmath}

\usepackage{graphicx}

\usepackage{CJKutf8}

\jvol{XX}
\jnum{XX}
\pubyear{2022}

\begin{document}

\sptitle{Microwave Magnetics}

\title{Temporal magnon-qubit Mach-Zehnder interferometer}

\author{Cody A. Trevillian\affilmark{1}*}
\author{Steven Louis\affilmark{2}*}
\author{Vasyl Tyberkevych\affilmark{1}*}

\affil{Department of Physics, Oakland University, Rochester, MI 48309, USA}
\affil{Department of Electrical and Computer Engineering, Oakland University, Rochester, MI 48309, USA}

\IEEEmember{*Member, IEEE}

\corresp{Corresponding authors:
Cody A. Trevillian (trevillian@oakland.edu),
Vasyl Tyberkevych (tyberkev@oakland.edu)
}

\markboth{Preparation of Papers for \emph{IEEE Magnetics Letters}}{Author Name}

\begin{abstract}
A temporal magnon-qubit Mach-Zehnder (MZ) interferometer is proposed.
The interferometer is based on controllable entanglement of a microwave qubit and a magnonic state, achieved by application of a pulsed magnetic field playing the role of a magnon-qubit temporal “beam splitter”.
Analogous to a typical MZ interferometer, the generated interference pattern of the final qubit population carries information about the magnon dynamics.
One important application of the proposed scheme is the study of single magnon decoherence.
Interestingly, this scheme allows one to independently determine rates of two possible decoherence channels.
This may help enable single magnon state applications and answer fundamental questions of quasi-particle decoherence at single quantum levels.
\end{abstract}

\begin{IEEEkeywords}
Microwave Magnetics, Magnonics, Magnetization Dynamics
\end{IEEEkeywords}

\maketitle

\section{INTRODUCTION}\label{introduction}

Mach-Zehnder interferometers (MZIs) have been a fundamental tool in physics for more than a century, originally designed to explore the interference patterns of light in the classical regime {[}Mach 1892; Zehnder 1891{]} and since evolving into versatile tools for exploring quantum phenomena {[}Caves 1981; DiDomenico 2004; Ekert 1998; Oliver 2005{]}.
In its simplest form, an MZI uses a pair of beam splitters to generate interference patterns by splitting a beam of light into two physical paths and later recombining them before measurement {[}Grigull 1967; Mach 1892; Wilkie 1963; Zehnder 1891{]}.
Continued development has shown that various excitations can be used to realize MZIs and MZI-like devices, including coherent light {[}Wilkie 1963{]}, lasers {[}Grigull 1967{]}, single photons {[}Rarity 1990; Ekert 1998; DiDomenico 2004; Xavier 2011{]}, electrons {[}Ji 2003{]}, phonons {[}Kaya 2014{]}, qubits {[}Oliver 2005{]}, and magnons {[}Balynsky 2017; Fetisov 1999; Khitun 2010; Kostylev 2005; Lee 2008; Rousseau 2015; Schneider 2008; Ustinov 2001{]}.
These recent extensions of MZIs illustrate the shift from classical to quantum regime explorations and highlight the adaptability of the underlying MZI framework to more complex quantum tasks.

One exciting extension of MZIs into the quantum regime is their application to hybrid quantum systems, where, among the various excitations used to realize MZIs, magnons—quanta of spin waves—hold the potential for unique functionality as they can easily couple with both bosonic and fermionic excitations {[}Awschalom 2021; Chumak 2022; Hou 2019; Huebl 2013; Li 2019, 2020; Tabuchi 2014, 2015; Xu 2021; Zhang 2014{]}.
Additionally, the resonant frequency of magnons can be tuned via external magnetic fields {[}Awschalom 2021; Hou 2019; Huebl 2013; Li 2019, 2020; Tabuchi 2014, 2015; Zhang 2014{]}, which means that a time-varying magnetic field may lead to novel functionality {[}Awschalom 2021; Chumak 2022; Xu 2021{]}.
This means that existing hybrid magnonic MZIs and MZI-like devices {[}Balynsky 2017; Fetisov 1999; Khitun 2010; Kostylev 2005; Lee 2008; Rousseau 2015; Schneider 2008; Ustinov 2001{]} could be developed further to study multiple different physical systems simultaneously.
This versatility makes magnons a compelling candidate for developing practical hybrid quantum systems and for investigating fundamental aspects of quantum dynamics at single-quanta level.

In this work, we propose to realize a temporal-domain MZI that operates in the quantum regime by entangling the states of a magnon and a qubit.
Specifically, this temporal magnon-qubit MZI leverages dynamic interaction between these two subsystems, where a pulsed magnetic field plays the role of a “beam splitter” for dynamically coupling and uncoupling the magnon and qubit modes, thereby controllably entangling and unentangling them.
The key difference between our approach and existing magnon-based MZI-like devices lies in the temporal nature of the interactions: instead of using spatially separated paths, we use time-dependent operations between the magnon and qubit subsystems to entangle and unentangle them.
This allows us to observe quantum interference effects in the final qubit state, which directly encodes information about the magnon dynamics during the evolution period.
This measurement scheme is especially well-suited for probing the decoherence processes of magnons, which are typically short-lived compared to qubits {[}Gurevich 1996{]}.

\begin{figure}[ht]
    \centering
    \includegraphics[scale=0.5]{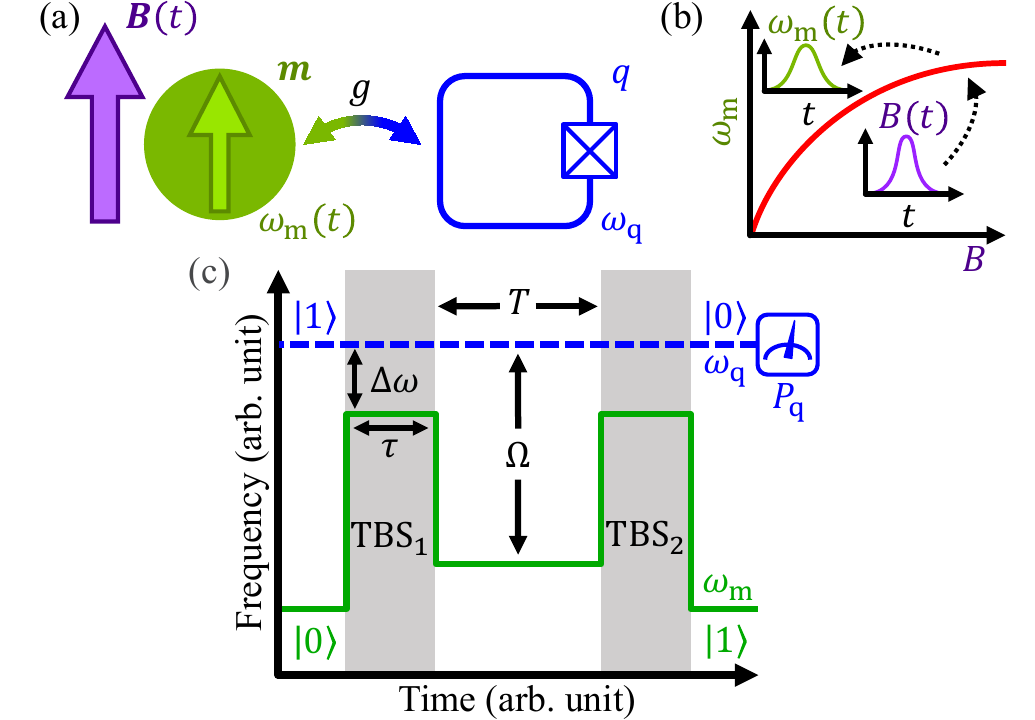} 
    \caption{
        Temporal magnon-qubit MZI measurement scheme and protocol.
        (a) Schematic diagram of a hybrid quantum system consisting of coupled magnon mode ($m$) and qubit ($q$).
        (b) Dynamic control of the magnon frequency $\omega_\mathrm{m}(t)$ by pulsed magnetic field $B(t)$.
        (c) Temporal MZI: two magnetic pulses $\mathrm{TBS}_1$ and $\mathrm{TBS}_2$ (shaded areas) serve as temporal “beam splitters” that couple magnon and qubit modes.
        In between the pulses, the two subsystems evolve independently.
        The interference result is measured as the final qubit state.
    }
    \label{fig:temporal_mzi}
\end{figure}

The need for this temporal magnon-qubit MZI is motivated by recent advances in quantum magnonics, including the development of single magnon sources {[}Chumak 2022{]} and strong coupling between magnons and other quantum systems {[}Awschalom 2021; Chumak 2022; Hou 2019; Huebl 2013; Li 2019, 2020; Tabuchi 2014, 2015; Xu 2021; Zhang 2014{]}.
These advances create a unique opportunity to study single magnon decoherence (SMD) mechanisms, which are critical for advancing quantum technologies that operate with finite magnon number states.
Our proposed interferometer provides a straightforward method for exploring these processes, particularly by enabling the independent measurement of different magnon decoherence channels, such as damping and dephasing.

The temporal magnon-qubit MZI introduced here provides a time-domain interferometric framework for probing single-magnon quantum dynamics in hybrid systems.
Although we focus on magnetic-field control of the magnon frequency, the interferometric protocol relies only on time-dependent control of the magnon-qubit detuning and can equivalently be implemented by tuning the qubit frequency.
In the following, we describe the scheme and analyze its operation under ideal and dissipative conditions.

\section{SCHEME OF TEMPORAL MZI}

We propose to realize a temporal magnon-qubit MZI using a simple hybrid quantum system shown schematically in Fig. \ref{fig:temporal_mzi}(a).
In this work we make no assumptions about the underlying geometry of such a system, and instead investigate what dynamics might be possible, agnostic to any implementation.
This system consists of a single magnon mode $\ket{m}$ with resonance frequency $\omega_\mathrm{m}$ and a single qubit $\ket{q}$ with resonance frequency $\omega_\mathrm{q}$.
The two subsystems are coupled with rate $g$.
Interaction between these two modes can be controlled by varying the magnon-qubit frequency gap $\Delta\omega = \omega_\mathrm{m} - \omega_\mathrm{q}$.
When $|\Delta\omega| \lesssim g$, a coherent exchange of quantum information between two subsystems is possible.
In the opposite limit $|\Delta\omega| \gg g$ the magnon and qubit modes evolve almost independently.

The magnon-qubit gap $\Delta\omega$ can be controlled dynamically by a time-dependent magnetic field $B(t)$ that changes the magnon frequency $\omega_\mathrm{m}(t)$ (see Fig. \ref{fig:temporal_mzi}(b)).
This means that by delivering short magnetic pulses one can controllably bring the two subsystems into resonance and the strong coupling regime ($|\Delta\omega| \ll g$), leading to entanglement between the magnon and qubit, or out of resonance, detuned from one another and into the weak coupling regime ($|\Delta\omega| \gg g$), causing the two modes to evolve independently.
These so-called \emph{dynamic magnon-qubit interactions} are the essential mechanism enabling the temporal interferometric scheme.

In the simplest case, we can consider a rectangular pulse of $\omega_m(t)$ with negligible rise and fall times (see $\mathrm{TBS}_1$ or $\mathrm{TBS}_2$ pulses in Fig. \ref{fig:temporal_mzi}(c)).
Then, the result of the dynamic magnon-qubit interaction can be described using only two parameters of the pulsed magnetic field profile, the interaction time $\tau$ and the frequency gap $\Delta\omega$.
This means that we can sequence multiple dynamic magnon-qubit interactions to perform operations analogous to beam-splitting and recombination in traditional MZIs.
In this temporal version, the “splitting” and “recombination” are achieved by applying two $\mathrm{TBS}$ pulses separated by a free evolution time $T$, during which the magnon and qubit evolve independently due to a large frequency gap $\Omega$.

The measurement scheme underlying the temporal magnon-qubit MZI follows a simple protocol, shown schematically in Fig. \ref{fig:temporal_mzi}(c).
Initially, the magnon and qubit subsystems are substantially detuned and the system is prepared in the unentangled state $\ket{1} \otimes \ket{0}$, with the qubit in the excited state $\ket{1}$ and the magnon mode in the ground state $\ket{0}$.
A magnetic field pulse $\mathrm{TBS}_1$ then acts as a magnon-qubit beamsplitter, reducing their detuning, temporarily coupling the two subsystems and allowing them to interact, thereby entangling their states.
Right after the end of the $\mathrm{TBS}_1$ the qubit-magnon system is in the entangled state $\alpha\ket{1} \otimes \ket{0} + \beta\ket{0} \otimes \ket{1}$, where the coefficients $\alpha$ and $\beta$ depend on the parameters of the “beamsplitter” pulse $\tau$ and $\Delta\omega$.
In particular, it is possible to choose $|\alpha| = |\beta| = 1/\sqrt{2}$, which corresponds to a usual 50:50 beamsplitter.

Following the $\mathrm{TBS}_1$ pulse, the magnon is detuned from the qubit to a frequency gap $\Omega$ and held fixed for a free evolution time $T$.
In practice, $|\Omega| \gtrsim 5g$ is sufficient to suppress coherent exchange such that the subsystems are effectively uncoupled from one another and undergo free evolution.
Please note that the system remains entangled during the free evolution time, but the subsystems evolve freely due to a large frequency gap $\Omega$. 

After the free evolution time, a second magnetic pulse $\mathrm{TBS}_2$ is applied.
The second $\mathrm{TBS}$ pulse is chosen to implement the inverse of the initial entangling operation, thereby coherently recombining the qubit and magnon amplitudes and effectively “unentangling” the subsystems.

Subsequent to  $\mathrm{TBS}_2$, the magnon and qubit subsystems return to a large detuning.
Finally, one measures the final qubit state, namely, probability $P_\mathrm{q}$ that the qubit is in the excited state.
These measurements allow one to recover the interference pattern formed by the qubit and magnon mode and reveal insights into the quantum processes happening during free magnon evolution (e.g., magnon decoherence mechanisms).

\section{METHODS}

\subsection{Hamiltonian dynamics without noise}

To simulate the evolution of the hybrid quantum system without noise, the Hamiltonian $\hat{\mathcal{H}}(t)$ for this system is written as
\begin{equation}
\hat{\mathcal{H}}(t)/\hbar = \omega_\mathrm{m}(t)\, \hat{m}^\dagger \hat{m} + \omega_\mathrm{q}\, \hat{q}^\dagger \hat{q} + g\, \hat{q}^\dagger \hat{m} + g^*\, \hat{m}^\dagger \hat{q},
\label{eq:Hamiltonian}
\end{equation}
where $\hat{m}^\dagger$ ($\hat{m}$) and $\hat{q}^\dagger$ ($\hat{q}$) are the creation (annihilation) operators for the magnon and qubit modes, respectively.
The qubit mode $\ket{q}$ is a fermionic quantum object described by Fock states $\ket{0}$ and $\ket{1}$.
The magnon mode is a bosonic quantum object described by Fock states $\ket{m}$ with integer non-negative number of magnons $m$.

The phase space of the hybrid quantum system is spanned by the tensor product states $\ket{q, m} = \ket{q} \otimes \ket{m}$.
Note that the Hamiltonian~(\ref{eq:Hamiltonian}) conserves the total number of quanta $N = q + m$, which means that we will only have a maximum of $N = 1$ quanta throughout the evolution of the system, due to the initial excitation being in the qubit mode.
This means that the state of the system $\ket{\psi(t)}$ is given by
\begin{equation}
\ket{\psi(t)} = c_{0,0}(t)\ket{0,0} + c_{0,1}(t)\ket{0,1} + c_{1,0}(t)\ket{1,0},
\label{eq:state}
\end{equation}
and evolves according to $\partial\ket{\psi(t)}/\partial t = -(i/\hbar)\, \hat{\mathcal{H}}(t)\ket{\psi(t)}$ such that
\begin{equation}
\begin{pmatrix}
\dot{c}_{0,0} \\
\dot{c}_{0,1} \\
\dot{c}_{1,0}
\end{pmatrix}
= -i
\begin{pmatrix}
0 & 0 & 0 \\
0 & \omega_\mathrm{m}(t) & g^* \\
0 & g & \omega_\mathrm{q}
\end{pmatrix}
\begin{pmatrix}
c_{0,0} \\
c_{0,1} \\
c_{1,0}
\end{pmatrix}.
\label{eq:coupled_eqs}
\end{equation}

The general solution to such a matrix equation is given by the unitary propagator $\hat{\mathcal{U}}(t, t_0)$:
\begin{equation}
\ket{\psi(t)} = \hat{\mathcal{U}}(t, t_0)\ket{\psi(t_0)} = \mathcal{T} \exp\left( -\frac{i}{\hbar} \int_{t_0}^{t} \hat{\mathcal{H}}(s)\, ds \right) \ket{\psi(t_0)},
\label{eq:unitary}
\end{equation}
where $\mathcal{T}$ denotes the time-ordering operation.
The propagator $\hat{\mathcal{U}}(t, t_0)$ can be found numerically using, e.g., the Magnus expansion, or, when $\hat{\mathcal{H}}$ is constant, $\hat{\mathcal{U}}(t, t_0)$ reduces to
\begin{equation}
\hat{\mathcal{U}}(t, t_0) = \exp\left( (-i/\hbar)\, \hat{\mathcal{H}}[\Delta\omega]\, (t - t_0) \right) \equiv \hat{\mathcal{U}}_0(\Delta\omega, \tau),
\label{eq:U0}
\end{equation}
and the result of the evolution depends only on the frequency gap $\Delta\omega$ and interaction time $\tau=t - t_0$ of the rectangular pulse.
For a sequence of rectangular magnetic field pulses, the overall propagator can be found as the product of corresponding $\hat{\mathcal{U}}_0$ propagators.
Thus, the propagator for the whole MZI sequence shown in Fig. \ref{fig:temporal_mzi}(c) is
\begin{equation}
\hat{\mathcal{U}}_\mathrm{MZI} = \hat{\mathcal{U}}_0(
-\Delta\omega, \tau
)\, \hat{\mathcal{U}}_0(\Omega, T)\, \hat{\mathcal{U}}_0(\Delta\omega, \tau).
\label{eq:UMZI}
\end{equation}

The final qubit population $P_\mathrm{q}$ is equal to the $|c_{1,0}|^2$ coefficient of the final state.
Here, the population refers to the probability of finding the qubit in its excited state. 
Note that in the MZI scheme, $c_{1,0}$ equals the $((1,0),(1,0))$ element of the matrix $\hat{\mathcal{U}}_\mathrm{MZI}$.

\subsection{Lindblad dynamics with noise}

To simulate evolution of our system with noise, we use the Gorini–Kossakowski–Sudarshan–Lindblad (GKSL) quantum master equation that describes evolution of the density matrix $\rho(t)$ given by
\begin{equation}
\dot{\rho}(t) = -\frac{i}{\hbar}\, [\hat{\mathcal{H}}(t), \rho(t)] + \sum_k \Gamma_k\, \mathcal{D}[L_k](\rho(t)),
\label{eq:gksl}
\end{equation}
where the density matrix $\rho(t) = \ket{\psi(t)}\bra{\psi(t)}$ and $\rho_{ij,kl} = \ket{ij}\bra{kl}$, and the commutator $[A, B] = AB - BA$.
The dissipator $\mathcal{D}$ describes the influence of the noise terms described by the Lindblad jump operators $L_k$ and acts on some operator $\hat{\mathcal{O}}$ as
\begin{equation}
\mathcal{D}[L](\hat{\mathcal{O}}) = L\cdot\hat{\mathcal{O}}\cdot L^\dagger - (1/2)\{L^\dagger\cdot L, \hat{\mathcal{O}}\},
\label{eq:dissipator}
\end{equation}
where the anti-commutator $\{A, B\} = AB + BA$.
The dissipator is linear such that for some set of jump operators $L = L_1 + L_2$, we have $\mathcal{D}[L](\hat{\mathcal{O}}) = \mathcal{D}[L_1](\hat{\mathcal{O}}) + \mathcal{D}[L_2](\hat{\mathcal{O}})$.

\begin{figure}
    \centering
    \includegraphics[scale=0.5]{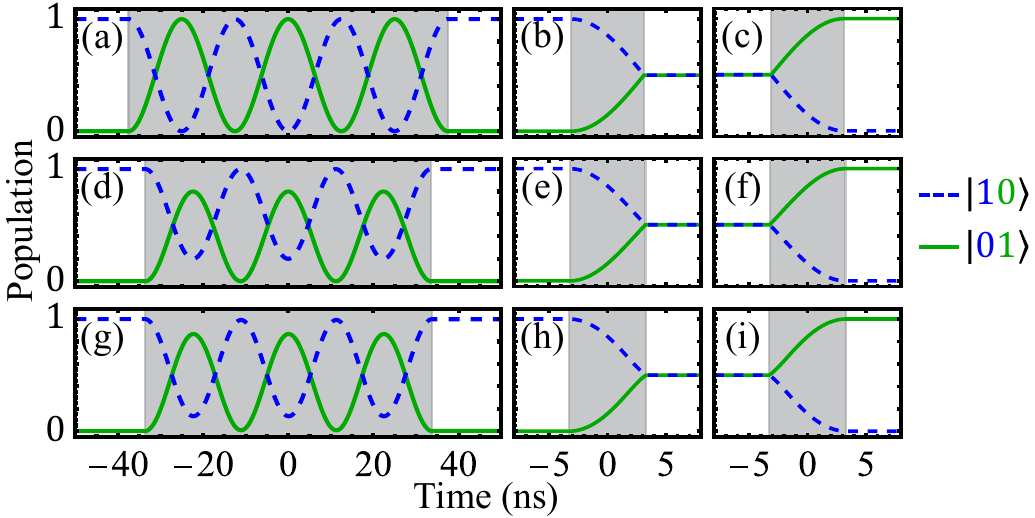}
    \caption{
    Operation of the magnon–qubit temporal beam splitter (TBS).
    Rows correspond to different detunings:
    (a, b, c) 
    $|\Delta\omega| = 0$,
    (d, e, f) 
    $|\Delta\omega| = g$,
    (g, h, i) 
    $|\Delta\omega| = g$ (with finite rise/fall time of 2.5~ns).
    Columns correspond to different operating regimes:
    (a, d, g) 
    coherent magnon–qubit dynamics during the coupling pulse,
    (b, e, h) 
    ``entangling'' operation for a balanced TBS pulse,
    (c, f, i) 
    corresponding ``unentangling'' operation.
    Balanced TBS pulse durations: 
    (b, c) $\tau \approx  6.25$~ns, 
    (e, f) $\tau \approx 6.49$~ns,
    (h, i) $\tau \approx 6.58$~ns.
    Magnon-qubit coupling rate $g = 2\pi \times 20 \mathrm{MHz}$.
    }
    \label{fig:tbs_operation}
\end{figure}

To numerically simulate the GKSL quantum master equation describing our system, we need to put Eq.~(\ref{eq:gksl}) into the superoperator or Liouville vectorized form given by
\begin{equation}
\dot{\vb*{\rho}} = \hat{\mathcal{L}}\cdot\vb*{\rho},
\label{eq:liouville}
\end{equation}
where $\vb*{\rho} = \mathrm{vec}(\rho)$ is a vectorized form of $\rho$, and $\hat{\mathcal{L}}$ is the superoperator associated with the generator of $\dot{\rho}$.
When $\hat{\mathcal{L}}$ is time-dependent,
\begin{equation}
\vb*{\rho}(t) = \mathcal{T} \exp\left( \int_{t_0}^{t} \hat{\mathcal{L}}(s)\, ds \right) \vb*{\rho}(t_0),
\label{eq:superoperator_evolution}
\end{equation}
which can be solved numerically using the Magnus expansion.
When $\hat{\mathcal{L}}$ is constant, the solution reduces to $\vb*{\rho}(t) = \exp(\hat{\mathcal{L}}\, \Delta t)\, \vb*{\rho}(t_0)$.
The final qubit population is given by $P_\mathrm{q} = \rho_{10,10}$ after evolution.

\section{RESULTS AND DISCUSSION}

\subsection{Magnon-qubit temporal beam splitter pulse}

The temporal beam splitter (TBS) between the magnon and qubit states is formed by resonantly coupling the two subsystems for some time $\tau$ and frequency gap $\Delta\omega$ allowing them to interact and coherently exchange quantum information (see Fig. \ref{fig:tbs_operation}(a, d, g)).
The scattering probability $P_\mathrm{TBS}$ between the states (the probability of an excitation scattering from the qubit to the magnon state) is then
\begin{equation}
P_\mathrm{TBS}(\Delta\omega, \tau) = \frac{(2g)^2}{\Delta\omega^2 + (2g)^2} \sin^2\left( \frac{\tau}{2} \sqrt{\Delta\omega^2 + (2g)^2} \right).
\label{eq:TBS}
\end{equation}
The functionality of the TBS pulse can also be described in terms of a splitting angle $\theta_\mathrm{BS}$ such that $P_\mathrm{TBS} = \sin^2(\theta_\mathrm{BS})$.
When $P_\mathrm{TBS}(\Delta\omega, \tau) = 1/2$, which corresponds to $\theta_\mathrm{BS} = \pi/4$, a balanced “50:50” magnon-qubit TBS pulse is achieved, as depicted.
Fig. \ref{fig:tbs_operation}(b, e, h; c, f, i) show that when the balanced TBS pulse is applied, it can maximally entangle the two subsystems or “unentangle” a previously entangled state, similar to a standard optical beamsplitter.
This serves as the primary interference mechanism enabling the functionality of the temporal magnon-qubit MZI.

Interestingly, a balanced TBS can be achieved for different detunings $\Delta\omega$ by properly adjusting the pulse duration $\tau$.
TBS pulses with different $\Delta\omega$ allow one to create entangled magnon-qubit states of the form $\left(\ket{0,1} + e^{i\phi} \ket{1,0}\right)/\sqrt{2}$ with different phase shift $\phi$.
For $|\Delta\omega| > 2g$, the maximum splitting angle $\theta_\mathrm{BS}$ falls below $\pi/4$ and a balanced TBS is no longer possible.

To justify the assumption of a rectangular pulse, we simulated a trapezoidal pulse in Fig.~\ref{fig:tbs_operation}(g, h, i) using finite rise and fall times of 2.5~ns, and large detuning before and after the pulse.
By comparing with the corresponding case in Fig.~\ref{fig:tbs_operation}(d, e, f), we found the balanced $\mathrm{TBS}$ pulse duration $\tau$ varied by less than $\sim100$~ps, thereby justifying the rectangular assumption.

In the following subsections, we utilize a nearly-resonant ($|\Delta\omega|=g$) balanced TBS pulse to further characterize the limiting factors of the temporal magnon-qubit MZI measurement scheme.
Then, we employ the scheme in exact resonance $\Delta\omega = 0$. 

\subsection{Influence of off-resonant coupling on free evolution}

\begin{figure}
    \centering
    \includegraphics[scale=0.5]{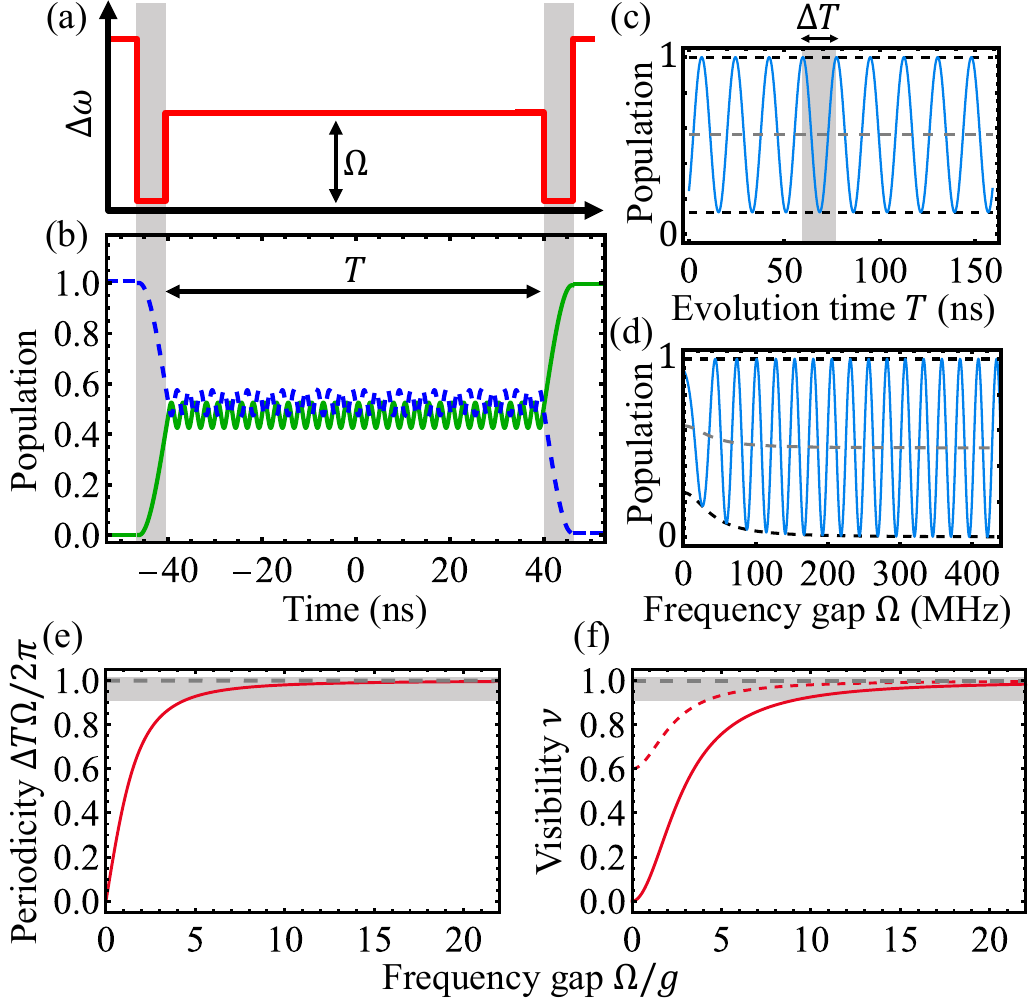} 
    \caption{
        Influence of the residual off-resonance coupling on temporal magnon-qubit MZI measurement scheme.
        (a) Cartoon diagram of magnon-qubit frequency gap $\Delta\omega(t)$ during MZI operation.
        (b) Numerical simulation of the magnon (green) and qubit (blue) populations during MZI operation.
        Fast small-amplitude oscillations during “free” evolution are due to the residual coupling.
        (c) Interference pattern for fixed-frequency-gap mode (fixed $\Omega$ and varying $T$).
        (d) Interference pattern for fixed-evolution-time mode (fixed $T$ and varying $\Omega$).
        (e) Dependence of the periodicity of the interference pattern on the frequency gap $\Omega$; gray highlight indicates periodicity $\geq$90\%.
        (f) Dependence of the visibility on the frequency gap $\Omega$; gray highlight indicates visibility $\geq$90\%, dashed and solid red lines denote TBS pulse detunings of $g$ and $2g$, respectively.
        Simulation parameters: $g = 2\pi \times 20$~MHz, $|\Delta\omega| = 2\pi \times 20$~MHz, $\tau \approx 6.49$~ns,
        (b) $\Omega = 2\pi \times 400$~MHz, $T \approx 79.3$~ns,
        (c) $\Omega = 2\pi \times 40$~MHz,
        (d) $T \approx 39.8$~ns.
    }
    \label{fig:residual_coupling}
\end{figure}

A key component of the proposed temporal magnon-qubit MZI measurement scheme is the interval of free evolution, the region of time during which magnon and qubit subsystems evolve independently.
The interference pattern generated by the final qubit population $P_\mathrm{q}$ thus relies on the parameters that govern this period of free evolution, the frequency gap $\Omega$ and the interaction time $T$.
Of these two parameters, the frequency gap $\Omega$ is the most critical, as it dictates the level to which the evolution is “free” or not, i.e., how much influence the residual off-resonance magnon-qubit coupling has on the dynamics of the system.
An example of such influence is shown in Fig. \ref{fig:residual_coupling}(b), where one can see small-amplitude high-frequency oscillations during the period of “free” magnon and qubit evolution.
A detuning of $\Omega \gtrsim 5g$ is sufficient to suppress off-resonant oscillations.

The main effects of the off-resonant coupling are the change of the period $\Delta T$ of the interference pattern and reduction of its visibility $\nu$.
The period $\Delta T$ corresponds to the regular fringe spacing of the interference pattern that converges to $2\pi/|\Omega|$ in the ideal case of a resonant $\mathrm{TBS}$ with $\Delta\omega=0$ and ``free'' evolution.
In the fixed-frequency-gap operation mode, when the gap $\Omega$ is kept constant and the free evolution time $T$ is varied, the period $\Delta T$ of the interference pattern is
$\Delta T = 2\pi/\sqrt{\Omega^2 + 4g^2} \approx 2\pi/|\Omega| - (4\pi g^2)/|\Omega|^3$,
and its deviation from the ideal case $\Delta T = 2\pi/|\Omega|$ is negligible for $|\Omega| > 5g$ (see Fig. \ref{fig:residual_coupling}(e)).

The visibility $\nu$ of the interference pattern is defined by the extrema of the qubit population,
\begin{equation}
\nu = (P_\mathrm{max} - P_\mathrm{min}) / (P_\mathrm{max} + P_\mathrm{min}).
\label{eq:visibility}
\end{equation}
Residual off-resonant coupling does not reduce visibility $\nu$ for resonant TBS $\Delta\omega = 0$ and unitary “free” magnon evolution.
However, for $\Delta\omega \neq 0$ or in the presence of non-unitary jump processes, the interference pattern visibility degrades at small frequency gaps $\Omega$ (see Fig. \ref{fig:residual_coupling}(f)).
Similar to the deviations in period $\Delta T$, these spurious effects are negligible for $|\Omega| > 5g$.
Thus, this inequality determines the condition under which magnon-qubit dynamics can be considered as uncoupled in the MZI scheme.

\subsection{Measurement of single magnon decoherence}

\begin{figure}
    \centering
    \includegraphics[scale=0.5]{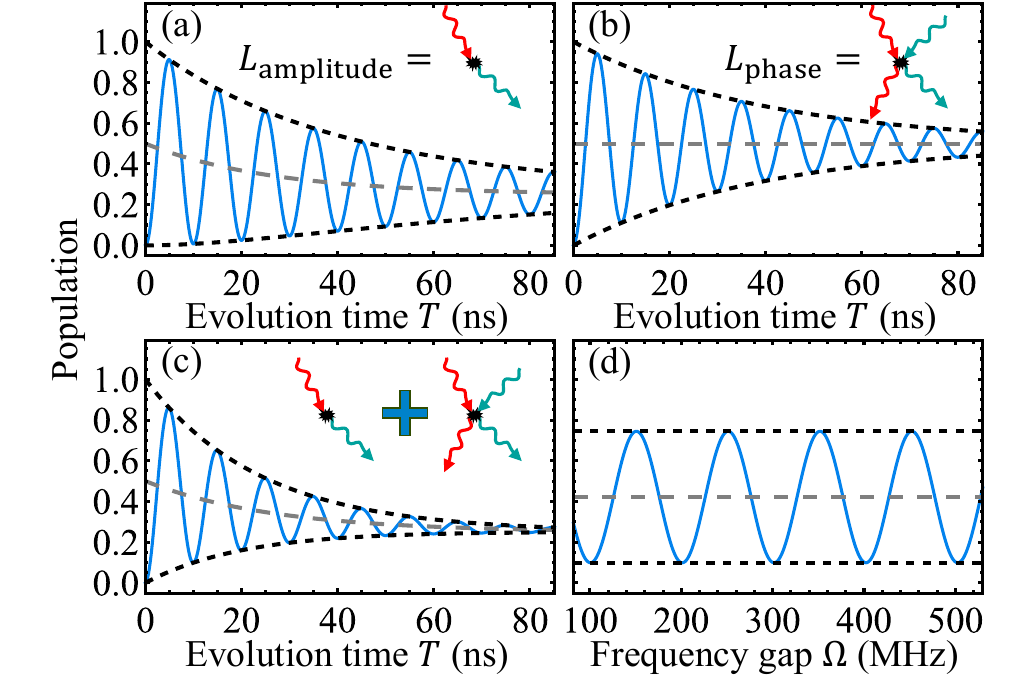} 
    \caption{
        Interference patterns for measurement of single magnon decoherence.
        (a–c) Fixed-frequency-gap interference pattern, (d) fixed-evolution-time interference pattern.
        (a) Amplitude noise, (b) phase noise, (c, d) mixture of amplitude and phase noise.
        Dashed black and gray lines in (a–d) indicate the envelope (i.e., the maxima $P_\mathrm{max}$ and minima $P_\mathrm{min}$) and average $P_\mathrm{avg}$ of the final qubit population (solid blue lines), respectively.
        Simulation parameters:
        (a–d) $g = 2\pi \times 20$~MHz, $\Delta\omega = 0$, $\tau \approx 6.25$~ns,
        (a–c) $\Omega = 2\pi \times 100$~MHz,
        (d) $T \approx 9.95$~ns,
        (a, c, d) $\Gamma_\mathrm{amplitude} \approx 37.7~\mu\mathrm{s}^{-1}$,
        (b–d) $\Gamma_\mathrm{phase} \approx 50.3~\mu\mathrm{s}^{-1}$.
    }
    \label{fig:smd_patterns}
\end{figure}

One interesting application of the proposed temporal magnon-qubit MZI is the measurement of single-magnon decoherence (SMD) mechanisms.
Here, we consider two potential noise channels that correspond to the more commonly encountered interactions that influence decoherence, namely, amplitude $L_\mathrm{amplitude}$ and phase $L_\mathrm{phase}$ noise.
Amplitude noise can be thought of as the process of a magnon encountering an inhomogeneity and scattering into a phonon.
In this case, the noise process is not number-conserving and can be described by the operator $L_\mathrm{amplitude} = \hat{m}$, with an accompanying amplitude noise rate $\Gamma_\mathrm{amplitude}$.
Phase noise can be thought of as the process of a magnon-phonon scattering, which results in a random change of the magnon phase.
In this case, the noise process is number-conserving and described by $L_\mathrm{phase} = \hat{m}^\dagger \hat{m}$, with an accompanying phase noise rate $\Gamma_\mathrm{phase}$.

The impact of these two SMD mechanisms can clearly be seen in Fig. \ref{fig:smd_patterns}.
When only $L_\mathrm{amplitude}$ is present as shown in Fig. \ref{fig:smd_patterns}(a), the average qubit population $P_\mathrm{avg}$ gradually decreases as the interaction time is increased, while $\nu$ remains relatively high.
In contrast, when only $L_\mathrm{phase}$ is present as shown in Fig. \ref{fig:smd_patterns}(b), $P_\mathrm{avg}$ remains unchanged at the 50\% value, but $\nu$ degrades with $T$.
In a system with a mixture of two noise channels, such as that shown in Fig. \ref{fig:smd_patterns}(c), the resulting interference pattern demonstrates both these effects.
It is interesting that both $\Gamma_\mathrm{amplitude}$ and $\Gamma_\mathrm{phase}$ rates can be found from the measurements of the MZI interference pattern.
Namely,
\begin{equation}
\Gamma_\mathrm{amplitude} = \frac{1}{T}\ln\left( \frac{1}{2(P_\mathrm{max} + P_\mathrm{min}) - 1} \right),
\label{eq:GammaAmplitude}
\end{equation}
and
\begin{equation}
\Gamma_\mathrm{phase} = \frac{1}{T}\ln\left( \frac{2(P_\mathrm{max} + P_\mathrm{min}) - 1}{(P_\mathrm{max} - P_\mathrm{min})^2} \right),
\label{eq:GammaPhase}
\end{equation}
where $P_\mathrm{max}$ and $P_\mathrm{min}$ are, respectively, the maximum and minimum final qubit populations.
In practical implementation of this SMD measurement scheme, it is better to work in the fixed-interaction-time ($T$) mode with varying frequency gap $\Omega$.
Because the decoherence processes depend only on the free-evolution time $T$, fixing $T$ results in an envelope that is independent of the frequency gap $\Omega$, while variations in $\Omega$ affect only the fringe periodicity. 
This mode simplifies measurements of $P_\mathrm{max}$ and $P_\mathrm{min}$, which are independent of the gap $\Omega$ (see Fig. \ref{fig:smd_patterns}(d)).






\begin{thebibliography}{}

\bibitem{bib1}
Awschalom D. D., Du C. R., He R., Heremans F. J., Hoffmann A., et.al. (2021), ``Quantum Engineering With Hybrid Magnonic Systems and Materials (Invited Paper),''
\emph{IEEE Trans. Quantum Eng.}, vol. 2, pp. 1–36, doi: 
\href{https://dx.doi.org/10.1109/TQE.2021.3057799}{10.1109/TQE.2021.3057799}.



\bibitem{bib2}
Balynsky M., Kozhevnikov A., Khivintsev Y., Bhowmick T., Gutierrez D., et.al. (2017), ``Magnonic interferometric switch for multi-valued logic circuits,''
\emph{J. Appl. Phys.}, vol. 121, no. 2, pp. 024504, doi: 
\href{https://dx.doi.org/10.1063/1.4973115}{10.1063/1.4973115}.

\bibitem{bib3}
Caves C. M. (1981), ``Quantum-mechanical noise in an interferometer,''
\emph{Phys. Rev. D}, vol. 23, no. 8, pp. 1693–1708, doi: 
\href{https://dx.doi.org/10.1103/PhysRevD.23.1693}{10.1103/PhysRevD.23.1693}.


\bibitem{bib4}
Chumak A. V., Kabos P., Wu M., Abert C., Adelmann C., Adeyeye A. O., Åkerman J., et.al.
(2022), ``Advances in Magnetics Roadmap on Spin-Wave Computing,''
\emph{IEEE Trans. Magn.}, vol. 58, no. 6, pp. 1–72, doi: 
\href{https://dx.doi.org/10.1109/TMAG.2022.3149664}{10.1109/TMAG.2022.3149664}.

\bibitem{bib5}
DiDomenico L. D., Lee H., Kok P., Dowling J. P. (2004), ``Quantum interferometric sensors,''
\emph{Quantum Sensing and Nanophotonic Devices}, vol. 5359, pp. 169–176.

\bibitem{bib6}
Ekert A. (1998), ``Quantum interferometers as quantum computers,''
\emph{Phys. Scr.}, vol. 1998, no. T76, pp. 218, doi: 
\href{https://dx.doi.org/10.1238/Physica.Topical.076a00218}{10.1238/Physica.Topical.076a00218}.

\bibitem{bib7}
Fetisov Y. K., Patton C. E. (1999), ``Microwave bistability in a magnetostatic wave interferometer with external feedback,''
\emph{IEEE Trans. Magn.}, vol. 35, no. 2, pp. 1024–1036, doi: 
\href{https://dx.doi.org/10.1109/20.748850}{10.1109/20.748850}.

\bibitem{bib8}
Grigull U., Rottenkolber H. (1967), ``Two-Beam Interferometer Using a Laser,''
\emph{JOSA}, vol. 57, no. 2, pp. 149–155, doi: 
\href{https://dx.doi.org/10.1364/JOSA.57.000149}{10.1364/JOSA.57.000149}.

\bibitem{bib9}
Gurevich A. G., Melkov G. A. (1996), ``Magnetization Oscillations and Waves,''
\emph{CRC Press, London}, doi: 
\href{https://dx.doi.org/10.1201/9780138748487}{10.1201/9780138748487}.

\bibitem{bib10}
Hou J. T., Liu L. (2019), ``Strong Coupling between Microwave Photons and Nanomagnet Magnons,''
\emph{Phys. Rev. Lett.}, vol. 123, no. 10, pp. 107702.

\bibitem{bib11}
Huebl H., Zollitsch C. W., Lotze J., Hocke F., Greifenstein M., et.al. (2013), ``High Cooperativity in Coupled Microwave Resonator Ferrimagnetic Insulator Hybrids,''
\emph{Phys. Rev. Lett.}, vol. 111, no. 12, pp. 127003.


\bibitem{bib12}
Ji Y., Chung Y., Sprinzak D., Heiblum M., Mahalu D., Shtrikman H. (2003), ``An electronic Mach–Zehnder interferometer,''
\emph{Nature}, vol. 422, no. 6930.

\bibitem{bib13}
Kaya O. A., Cicek A., Salman A., Ulug B. (2014), ``Acoustic Mach–Zehnder interferometer utilizing self-collimated beams in a two-dimensional phononic crystal,''
\emph{Sens. Actuators B Chem.}, vol. 203, pp. 197–203, doi: 
\href{https://dx.doi.org/10.1016/j.snb.2014.06.097}{10.1016/j.snb.2014.06.097}.

\bibitem{bib14}
Khitun A., Bao M., Wang K. L. (2010), ``Magnonic logic circuits,''
\emph{J. Phys. Appl. Phys.}, vol. 43, no. 26, pp. 264005, doi: 
\href{https://dx.doi.org/10.1088/0022-3727/43/26/264005}{10.1088/0022-3727/43/26/264005}.

\bibitem{bib15}
Kostylev M. P., Serga A. A., Schneider T., Leven B., Hillebrands B. (2005), ``Spin-wave logical gates,''
\emph{Appl. Phys. Lett.}, vol. 87, no. 15, pp. 153501.

\bibitem{bib16}
Lee K.-S., Kim S.-K. (2008), ``Conceptual design of spin wave logic gates based on a Mach–Zehnder-type spin wave interferometer for universal logic functions,''
\emph{J. Appl. Phys.}, vol. 104, no. 5, pp. 053909, doi: 
\href{https://dx.doi.org/10.1063/1.2975235}{10.1063/1.2975235}.

\bibitem{bib17}
Li Y., Polakovic T., Wang Y.L., Xu J., Lendinez S., et.al. (2019), ``Strong Coupling between Magnons and Microwave Photons in On-Chip Ferromagnet-Superconductor Thin-Film Devices,''
\emph{Phys. Rev. Lett.}, vol. 123, no. 10, pp. 107701.


\bibitem{bib18}
Li Y., Zhang W., Tyberkevych V., Kwok W.-K., Hoffmann A., Novosad V. (2020), ``Hybrid magnonics: Physics, circuits, and applications for coherent information processing,''
\emph{J. Appl. Phys.}, vol. 128, no. 13, pp. 130902, doi: 
\href{https://dx.doi.org/10.1063/5.0020277}{10.1063/5.0020277}.

\bibitem{bib19}
Mach L. (1892), ``Ueber einen interferenzrefraktor,''
\emph{Z. Für Instrumentenkunde}, vol. 12, pp. 89–93.

\bibitem{bib20}
Oliver W. D., Yu Y., Lee J. C., Berggren K. K., Levitov L. S., Orlando T. P. (2005), ``Mach-Zehnder Interferometry in a Strongly Driven Superconducting Qubit,''
\emph{Science}, vol. 310, no. 5754, pp. 1653–1657, doi: 
\href{https://dx.doi.org/10.1126/science.1119678}{10.1126/science.1119678}.

\bibitem{bib21}
Rarity J. G., Tapster P. R., Jakeman E., Larchuk T., Campos R. A., Teich M. C., Saleh B. E. A. (1990), ``Two-photon interference in a Mach-Zehnder interferometer,''
\emph{Phys. Rev. Lett.}, vol. 65, no. 11, pp. 1348–1351, doi: 
\href{https://dx.doi.org/10.1103/PhysRevLett.65.1348}{10.1103/PhysRevLett.65.1348}.

\bibitem{bib22}
Rousseau O., Rana B., Anami R., Yamada M., Miura K., Ogawa S., Otani Y. (2015), ``Realization of a micrometre-scale spin-wave interferometer,''
\emph{Sci. Rep.}, vol. 5, no.1.

\bibitem{bib23}
Schneider T., Serga A. A., Leven B., Hillebrands B., Stamps R. L., Kostylev M. P. (2008), ``Realization of spin-wave logic gates,''
\emph{Appl. Phys. Lett.}, vol. 92, no. 2, pp. 022505, doi: 
\href{https://dx.doi.org/10.1063/1.2834714}{10.1063/1.2834714}.

\bibitem{bib24}
Tabuchi Y., Ishino S., Ishikawa T., Yamazaki R., Usami K., Nakamura Y. (2014), ``Hybridizing Ferromagnetic Magnons and Microwave Photons in the Quantum Limit,''
\emph{Phys. Rev. Lett.}, vol. 113, no. 8, pp. 083603, doi: 
\href{https://dx.doi.org/10.1103/PhysRevLett.113.083603}{10.1103/PhysRevLett.113.083603}.

\bibitem{bib25}
Tabuchi Y., Ishino S., Noguchi A., Ishikawa T., Yamazaki R., Usami K., Nakamura Y. (2015), ``Coherent coupling between a ferromagnetic magnon and a superconducting qubit,''
\emph{Science}, vol. 349, no. 6246, pp. 405–408, doi: 
\href{https://dx.doi.org/10.1126/science.aaa3693}{10.1126/science.aaa3693}.

\bibitem{bib26}
Ustinov A. B., Kalinikos B. A. (2001), ``Nonlinear microwave spin wave interferometer,''
\emph{Tech. Phys. Lett.}, vol. 27, no. 5, pp. 403–405, doi: 
\href{https://dx.doi.org/10.1134/1.1376765}{10.1134/1.1376765}.

\bibitem{bib27}
Wilkie D., Fisher S. A. (1963), ``Measurement of Temperature by Mach-Zehnder Interferometry,''
\emph{Proc. Inst. Mech. Eng.}, vol. 178, no. 1, pp. 461–470.

\bibitem{bib28}
Xavier G. B., Weid J. P. von der (2011), ``Stable single-photon interference in a 1 km fiber-optic Mach–Zehnder interferometer with continuous phase adjustment,''
\emph{Opt. Lett.}, vol. 36, no. 10, pp. 1764–1766, doi: 
\href{https://dx.doi.org/10.1364/OL.36.001764}{10.1364/OL.36.001764}.

\bibitem{bib29}
Xu J., Zhong C., Han X., Jin D., Jiang L., Zhang X. (2021), ``Coherent Gate Operations in Hybrid Magnonics,''

\bibitem{bib30}
Zehnder L. (1891), ``Ein neuer interferenzrefraktor,''
\emph{Z. Für Instrumentenkunde}, vol. 11, pp. 275–285.

\bibitem{bib31}
Zhang X., Zou C.-L., Jiang L., Tang H. X. (2014), ``Strongly Coupled Magnons and Cavity Microwave Photons,''

\end{thebibliography}
\end{document}